\documentclass[twocolumn,aps,floatfix]{revtex4}
\usepackage{amsmath}
\usepackage{graphicx}
\usepackage{dcolumn}
\usepackage{bm}

\newcommand{\vev}[1]{\langle #1 \rangle}
\newcommand{\be}{\begin{equation}}
\newcommand{\ee}{\end{equation}}
\newcommand{\bea}{\begin{eqnarray}}
\newcommand{\eea}{\end{eqnarray}}
\newcommand{\mk}{{\mathbf k}}
\def\dup{\;\raise1.0pt\hbox{$'$}\hskip-6pt\partial\;}
\def\ddn{\;\overline{\raise1.0pt\hbox{$'$}\hskip-6pt\partial}\;}
\newcommand{\f}{\begin{equation}}
\newcommand{\ff}{\end{equation}}

\begin{document}


\title{Anomalous CMB polarization and gravitational chirality}

\author{Carlo R. Contaldi$^1$, Jo\~ao Magueijo$^1$ and Lee Smolin$^2$}%
\affiliation{%
$^1$Theoretical Physics, Imperial College, Prince
 Consort Road, London, SW7 2BZ, U.K.}%
\affiliation{$^2$Perimeter Institute, 31 Caroline St N, Waterloo, ON N2L 2Y5,
Canada}

\date{\today}

\begin{abstract}
We consider the possibility that gravity breaks parity, with left and
right handed gravitons coupling to matter with a
different Newton's constant and show that this would affect their zero-point
vacuum fluctuations during inflation. Should there
  be a cosmic background of gravity waves,  the effect would translate into
  anomalous CMB polarization. Non-vanishing $TB$ (and $EB$)
  polarization components emerge, revealing interesting experimental
  targets. Indeed if reasonable chirality is present a TB 
measurement would provide the easiest way  to detect
a gravitational wave background. We speculate on the theoretical
implications of such an observation.  
\end{abstract}

\pacs{Valid PACS appear here}
\maketitle

{\it Introduction} The fact that the standard model is chiral, allied
with the belief that all the forces of nature should be unified, leads to the natural suggestion that 
 gravity itself might be chiral~\cite{steph,lee}.
Furthermore, the formulation of relativity due to Cartan and
Kibble~\cite{kibble}, as well as its development in the Ashtekhar
formalism~\cite{rovelli}, shows that gravity has a definite propensity
for chirality due to the presence of  terms in the action exhibiting odd
parity. The extent of parity violation, however, remains
unresolved, both at the classical and quantum
levels~\cite{freidel,Randono,rovellipar}.  In this letter we explore
some implications of gravitational chirality that would improve the
prospects of gravitational wave detection in CMB experiments.

Following the successes in CMB temperature anisotropy mapping, the
future of CMB physics now lies in improving polarization
measurements~\cite{zaldarriaga,polrev1,polrev2}.  Polarization can
be decomposed into electric ($E$) and magnetic ($B$) modes, with positive
and negative parities respectively.  Correlators between these modes
and with the temperature ($T$) may be obtained, and in the absence of
parity violation the only non-vanishing quadratic correlators are $TT$,
$TE$, $EE$ and $BB$.  Scalar perturbations (i.e. density fluctuations) are
known to seed $E$-mode polarization only, and so affect $TT$, $TE$ and $EE$
correlators.  Tensor modes, the hallmark of a gravitational wave
background, are needed in order to generate $B$-mode polarization. For
this reason it has been suggested that a $BB$ measurement would be a
choice method for a first detection of gravitational waves. The
effect, however, is predicted to be very small, presenting a major
experimental challenge, particularly when galactic foregrounds are
considered.

But what if parity is violated by gravity? Then one could expect a
non-vanishing $TB$ correlator.  This correlation may provide the
easiest way to detect $B$-mode polarization---and by implication
gravitational waves---for the same reason that $TE$ correlations are
easier to measure than $EE$ ones \cite{wmap1pol}: 
they correlate something big ($T$)
with something small ($E$ or $B$), rather than two smaller quantities
($EE$ or $BB$). 
The proposed $TB$ measurement means nothing short of catching two
pigeons with one stone.  Should gravity be
chiral {\it and} should there be a gravitational wave background it
would be easier to detect them together, via their combined effects,
rather than separately.  The catch: if at least one of these two
premises is violated then no effect is expected. Either premise on its
own would not lead to $TB$, and a lack of $TB$ measurement would not
disprove either.  What is at stake, however, is of such importance
that we believe the issue deserves to be investigated further,
experimentally and theoretically. We speculate towards the end of
the paper on the theoretical implications of gravitational parity violation.

{\it Parity breaking in linearized gravity and the implications for inflation}
We first consider the implications of chiral symmetry breaking 
in {\it linearized } gravity for the production
of gravitational waves during inflation. We shall show that a 
chiral imprint would be left in the gravitational wave background,
with dramatic implications for CMB polarization.
Consider a metric of the form $g_{\mu\nu}=a^2(\eta_{\mu\nu}+h_{\mu\nu})$. 
Usually the second order action may be written as 
\begin{equation}\label{haction}
S=\frac {1}{64 \pi G}\int a^2 (\dot h^{ij}\dot h_{ij} - h^{ij,k}
h_{ij,k}) \; d^3x,
\end{equation}
and with expansion
\begin{equation}
h_{ij}=\int \frac{d^3k}{(2\pi)^3 2 k^0}\sum _{r=L,R} 
A^r(\mathbf k) \epsilon _{ij}^r h(k,\eta) e^{i{\mathbf k}\cdot{\mathbf x}}
+ h.c. \, ,
\end{equation}
the action and Hamiltonian break into left and right components, each
real on its own. Circular polarization states in a frame aligned 
with $\mathbf k$ have tensors:
\begin{eqnarray} 
\epsilon^{R/L}_{ij}&=&\frac{1}{\sqrt2} \left(\begin{array}{ccc}
0&0&0\\
0&1&\pm i\\
0&\pm i&-1\end{array}\right),
\end{eqnarray}
and the split in $S$ can be traced to the orthonormality conditions
for these tensors.

For $k\eta\ll 1$ we choose boundary condition
$h\rightarrow e^{-i\omega \eta}$, but more generally in 
an expanding universe we have equation:
\begin{equation}\label{veqn}
v''+{\left(k^2-\frac{a''}{a}\right)} v=0,
\end{equation}
with $v(k)\propto a h$. 
The usual calculation of inflationary vacuum quantum fluctuations
relies on the fact that the action becomes a regular scalar field action 
with normalizations:
$v={a h}/{\sqrt {32 \pi G}}$.
Canonical quantization inside the horizon supplies a
vacuum fluctuation that can then be followed outside the horizon
with the textbook result \cite{mukh,liddle}. 

There is nothing in the linearized theory
that prevents us from attributing a different gravitational constant 
to R and L gravitons. They are genuinely independent degrees 
of freedom and (\ref{haction}) could be replaced by
\begin{equation}
S=\frac{s^R}{64\pi G^R}+\frac{s^L}{64\pi G^L},
\end{equation}
leading to a Hamiltonian:
\begin{equation}
H=\frac{1}{64\pi}\int \frac{d^3 k}{(2\pi)^3 2k^0} \sum_{r=L,R}
\frac{k^0}{G_r}{\left(A^\dagger_r({\mathbf k})A_r({\mathbf k})+
\frac{1}{2}\right)}.
\end{equation}
Canonical quantization can be studied as usual and a vacuum
fluctuation found, from the boundary condition inside
the horizon, and followed until it freezes out. We find
the straightforward modification of the standard result:
\begin{equation}\label{eq:power}
P_{R/L}(k)k^3=\frac{4 G^{R/L}}{\pi}H^2_{|k=aH},
\end{equation}
which is scale-invariant when the background is close to de~Sitter
with $H$ constant.  A crucial modification in the normalizations,
however, slips into the CMB polarization calculation. For reasons 
to be made obvious later, we shall parametrize this asymmetry by:
\begin{equation}\label{Gform}
G^{R/L}=\frac{G}{1\mp \frac{1}{\gamma}}
\end{equation}
A large $|\gamma|$ means no measurable chirality. The sign of $\gamma$ 
matters and $\gamma>0$ means stronger gravity for R gravitons.
If $|\gamma|<1$ then gravity becomes repulsive for one of the
R/L modes. For $|\gamma|=1$ the coupling constant for one of 
the modes diverges and we can ignore the other mode.

It is important to note that for linear polarization the basis change
induced by the transformation ${\mathbf k} \rightarrow -\mathbf k$ is
such that $\epsilon^+_{ij}(-\mk) =\epsilon_{ij}^+(\mk)$ and
$\epsilon^\times_{ij}(-\mk) =-\epsilon_{ij}^\times(\mk)$. This results
in reality conditions $h_+(\mk)=h_+^\star(-\mk)$ and
$h_\times(\mk)=-h_\times^\star(-\mk)$. Using the relation 
$h_{R/L}=({h^+\mp i h^\times})/{\sqrt 2}$,
this implies the separate reality conditions
$h_R(\mk)=h_R^{\star}(-\mk)$ and 
$h_L(\mk)=h_L^\star(-\mk)$
In the inflationary setting we have described we therefore have:
\begin{eqnarray}\label{eq:prpl}
{\langle h_R(\mk)h^\star_R(\mk ')\rangle}&=&\delta(\mk-\mk')P_R(k),\nonumber\\
{\langle h_L(\mk)h^\star_L(\mk ')\rangle}&=&\delta(\mk-\mk')P_L(k),\nonumber\\
{\langle h_R(\mk)h^\star_L(\mk ')\rangle}&=&0
\end{eqnarray}
and 
obviously $P_R(k)\neq P_L(k)$ does not contradict the reality conditions. 


{\it CMB polarization in chiral gravity}
We now examine the impact of such chirality upon the CMB.
Linear polarization of the
radiation is described by the three Stokes parameters $I$, $Q$, and
$U$. The $I$ component is invariant under right handed rotations
$\psi$ about the line of sight vector ${\bf \hat n}$ while $Q$ and $U$
components transform as $Q'=Q\cos 2\psi + U\sin 2\psi$ and $U'=-Q\sin
2\psi + U\cos 2\psi$. We can construct two fields with spin-2 symmetry
on the sky from the $Q$ and $U$ parameters and rotate as
 $(Q\pm iU)'({\bf\hat n})= e^{\mp2i\psi}(Q\pm iU)({\bf\hat n})$.
On the sky the spin-2 fields can be decomposed onto the basis of
spin-2 weighted spherical harmonic functions $_{\pm 2}Y_{\ell m}({\bf
\hat n})$ as
$ (Q\pm iU)({\bf\hat n}) = \sum_{\ell m} a_{\pm 2, \ell m} \,{}_{\pm
2}Y_{\ell m}({\bf
\hat n})$.
Polarization induced by Thompson scattering of a plane wave
perturbation is best considered in the local frame ${\bf \hat
  e}_1,{\bf \hat e}_2,{\bf \hat e}_3 $ whose axis ${\bf \hat e}_3$ is
aligned with plane wave vector ${\bf k}$
\cite{polnarev,kosowsky}. In the aligned frame only the
$Q$ Stokes parameter is generated and its magnitude is proportional
to $(1-\mu^2)e^{\pm im\phi}$ where $\mu$ is
the angle between the plane wave and the outgoing photon momentum $\mu
= {\bf \hat k}\cdot {\bf \hat n}$, $\phi$ is the azimuthal angle of
the wavevector and $m =0, \,1,\, 2$ for scalar, vector and tensor
sources respectively. The induced polarization can then be
rotated and averaged over the whole sky to obtain the tensor generated
spin-2 fields \cite{polnarev} e.g. for the tensor sources;
\begin{eqnarray}
  (Q\pm iU)({\bf \hat n},{\bf k}) = \left[(1\mp \mu)^2e^{i2\phi}h_R({\bf
      k}) \right.+ \hspace{1.3cm} \nonumber\\
    \left.(1\pm \mu)^2e^{-i2\phi}h_L({\bf
      k}) \right]\Delta_P(\mu,k),
\end{eqnarray}
where $\Delta_P(\mu,k)$ are the tensor polarization source functions
for the  perturbation to the photon phase space density
\cite{bond} and are obtained as solutions of the full
Einstein-Boltzmann system.

It is convenient to define two rotationally invariant, spin-0 fields
by acting twice on the spin-2 fields with spin raising and lowering
operators $\dup$ and $\ddn$ \cite{zaldarriaga}
\begin{eqnarray}
  E({\bf \hat n}) &=& - \frac{1}{2}\left[ \dup^2(Q+ iU)({\bf \hat n})
  + \ddn^2(Q- iU)({\bf \hat n})\right],\nonumber\\
 B({\bf \hat n}) &=& \frac{i}{2}\left[ \ddn^2(Q+ iU)({\bf \hat n})
  - \dup^2(Q- iU)({\bf \hat n})\right].
\end{eqnarray}
The two rotationally invariant fields have opposite parity with
respect to fields with opposite spin. Scalar perturbations give
$\dup^2(Q+ iU) = \ddn^2(Q- iU)$ since the the spin-2 fields generated
have no parity dependence. Thus scalars do not source the $B$
field. Tensors however generate the parity sensitive $B$-field due to
the extra $e^{\pm i2\phi}$ dependence.

Solving for the $E$, $B$ and $T$ fields at late time and expanding
onto spherical harmonic coefficients $a_{\ell m}^{T,E,B}$ one can define
the present day angular power spectra $C^{XY}_\ell =1/(2\ell+1)\sum_m
\vev{a^X_{\ell m}a^{Y\, \star}_{\ell m}} $ for all correlations of the
three fields. Here we list only the cross-correlation spectra which
are of particular interest
\begin{eqnarray}\label{eq:spectra}
  C^{TE}_\ell &=& 8\pi \int d k
P^+(k)\Delta^T_\ell(k,\eta_0)\Delta^E_\ell(k',\eta_0),\nonumber\\
  C^{TB}_\ell &=& 8\pi \int d k
P^{-}(k)\Delta^T_\ell(k,\eta_0)\Delta^B_\ell(k',\eta_0),\nonumber\\
  C^{EB}_\ell &=& 8\pi \int dk
P^{-}(k)\Delta^E_\ell(k,\eta_0)\Delta^B_\ell(k',\eta_0),
\end{eqnarray}
where the $\Delta^X_\ell(k,\eta_0)$ are the Legendre expanded
radiation transfer functions integrated to the present time. The
functions $P^+(k)=P_R(k)+P_L(k)$ and $P^-(k)=P_R(k)-P_L(k)$ are the sum and
difference of the $R$
and $L$ mode power spectra (\ref{eq:prpl}) under the assumption of isotropy.
Following
(\ref{eq:power}) we can write
\begin{equation}
  P^+(k) =  \frac{P_h(k)}{\left(1-\frac{1}{\gamma2}\right)}\ \ \mbox{and}
  \ \ P^-(k) = \frac{P_h(k)}{2\gamma\left(1-\frac{1}{\gamma2}\right)},
\end{equation}
where $P_h(k)$ is a reference spectrum for the combination of the two
gravitational modes for the standard case ($\gamma\rightarrow
\infty$). As shown in (\ref{eq:spectra}) any tensor contribution to
the $TB$ and $EB$ cross-correlation spectra vanishes for the standard
parity invariant case. Thus any non-zero $TB$ and $EB$ signal would be
an unambiguous indication of new parity breaking physics either in the
primordial gravitational wave spectrum \cite{kamionkowskitb} or from
effects along the line of sight that rotate the polarizations
\cite{Xia}.

\begin{figure}[t]
\centering
\includegraphics[width=9cm,angle=0]{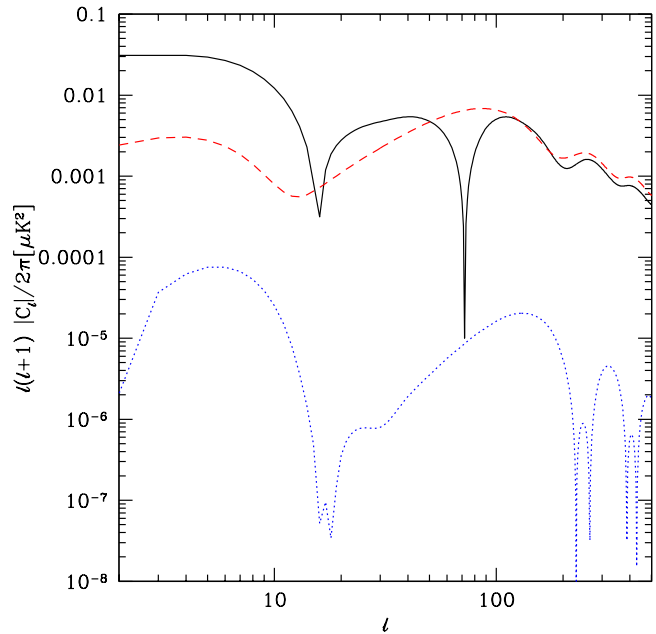}
\caption{Tensor contribution to the $TB$ (solid, black), $BB$ (dashed,
red), and $EB$ (dotted, blue) spectra for a standard $\Lambda$CDM
model with tensor to scalar ratio $r=0.1$ and chirality 
parameter $\gamma = 10$.}\label{fig:spectra}
\end{figure}

{\it Results} Standard line of sight, Einstein-Boltzmann codes (e.g. CAMB
\cite{lewis}\footnote{http://camb.info}) can be easily modified to include
the calculation of the
$TB$ and $EB$ spectra and in Fig.~\ref{fig:spectra} we show the tensor
sensitive combinations obtained for a model with $\gamma=10$ and
tensor to scalar ratio $r=0.1$.
Searching for such a unique signal in the cross-correlation spectra
offers some observational advantages. As mentioned previously the $TB$
signal is larger than the pure $BB$ correlation but also does not suffer
from
noise bias in the absence of noise correlations between total
intensity and polarization sensitive measurements. In addition the
$TB$ spectrum is free of any ambiguities induced by the coupling of
$E$ and $B$-modes due to cut--sky effects in multipole space.

Observationally, the strength of the effect is determined by both the
amplitude of the gravitational wave background, usually denoted by the
ratio of primordial tensor-to-scalar spectra normalizations
$r=A_h/A_S$, and the value of our parity breaking measure
$\gamma$. The ratio of quadrupole power of the two, opposite parity
tensor contributions can be approximated as $C_2^{TB}/C_2^{BB}\approx
\alpha_2/\gamma$, where $\alpha_2$ is a depends on the exact
cosmology and $\alpha_2\sim 200$ for a standard $\Lambda$CDM model. In
this case the $TB$ contribution will be larger than the $BB$ one for
$\gamma < 102$. Alternatively we can examine the overall amplitude of
the effect by comparing to the scalar contribution to the total
intensity spectrum
\begin{equation}
  \frac{C_2^{TB}}{C_2^{TT(S)}} \approx \beta_2\,
\frac{r}{\gamma}\frac{1}{(1-\frac{1}{\gamma2})} \sim 1\times 10^{-3} \,
\frac{r}{\gamma},
\end{equation}
for $\gamma \gg 1$ and where $\beta_2\sim 1\times 10^{-3}$ is again a
reference value for a
standard $\Lambda$CDM model.

CMB results have not yet reached the sensitivity required to impose
interesting limits but most polarisation experiments are now reporting
the parity violating spectra $TB$ and $EB$ in addition to the usual
four since these also provide useful consistency checks on
instrumental and analysis methods. The best constraint so far are from
the latest WMAP 5-year results \cite{wmap5}
\footnote{http://lambda.gsfc.nasa.gov} which observed a $TB$
quadrupole $\ell(\ell+1)C_\ell^{TB}/2\pi \approx 1.26 \pm 0.87 \mu K^2$. This
can be interpreted broadly as a 3-$\sigma$ upper bound of $-1.5 < C_2^{TB} <
4 \mu K^2$ which  translates into a limit of $\gamma-1> 0.4\,r$
and $\gamma+1<- 0.15\,r$ (where for simplicity we 
ignored the $|\gamma|<1$ possibility). 
We are still in the regime $|\gamma| \rightarrow
1$, but
future data will give much more stringent constraints of $\gamma
\gg 1$; or else provide a detection.

{\it Motivating chiral gravity}
What would be the theoretical implications of such an observation?
While linearized gravity is all that is needed to deduce a spectrum for tensor fluctuations during inflation, it is generally assumed that that theory is a linearization of a classical non-linear gravitational theory, which is general relativity or a closely related modification.  General relativity is parity symmetric, so it is pertinent to ask how radical the modifications of its principles need be to allow parity asymmetry in the form of $G_L \neq G_R$.

Chiral gravitation has been associated with a Chern-Simons 
term~\cite{cs,Xia,kamionkowskitb} coupled to a dilaton, or the presence of 
spinors~\cite{rovelli,freidel,Randono}.  But none of these mechanisms induce parity breaking at leading order in the graviton propagator, as is implied by $G_L \neq G_R$.   But the possibility of such a leading order effect  can be motivated
from several considerations including Euclidean gravity and the fact that $CP$ violating instanton effects are expected to arise  in a path integral quantization of chiral actions such as the JSS and Plebanski actions~\cite{Soo}. Note also that in the linearized calculation presented
above all that may be needed from the full theory is parity violation 
in the action, as opposed to the field equations, since we're only
concerned with the {\it quantum} zero-point fluctuations.
Several
actions in use, such as the $JSS$ and Holtz actions, already have this property.

We {\it sketch} how this may come about, leaving details to~\cite{future}. 
Let us consider the Euclidean action:
\begin{equation}\label{holst}
S=\frac{1}{32\pi G}\int {\left( \epsilon_{abcd}
e^a\wedge e^b \wedge R^{cd}
-\frac{2}{\gamma} e_a\wedge e_b\wedge R^{ab}\right)},
\end{equation}
where $e_a$ is the tetrad, $R_{ab}$ the curvature
and $\gamma$ is the Immirzi parameter
(we'll use latins for the SO(4) group index).
Introducing the area form $\Sigma_{ab}=e_a\wedge e_b$ 
the action can be written in terms of 2-forms as 
\begin{equation}
S=\frac{1}{16\pi G}\int \Sigma^{ab}\wedge{\left(\star R_{ab}-\frac{1}{\gamma}
R_{ab}\right)}.
\end{equation}
where $\star$ represents the dual form.
Splitting into self-dual and anti-self-dual components,
we have $S=S^++S^-$ with the illuminating result
\begin{equation}\label{plebs}
S^\pm=\frac{1}{16 \pi G}\int \Sigma^{ab\pm}\wedge R^\pm_{ab}
{\left(\pm 1-\frac{1}{\gamma}\right)},
\end{equation}
(here $F^\pm=(F\pm \star F)/{2}$ and 
$\star F^{\pm}=\pm F^\pm$). 
Thus, if $\gamma$ is real, we find a shift in 
the gravitational constant for + and -
with a formula identical to (\ref{Gform}) used in our phenomenology. 
It is tempting to translate this argument into SO(3,1) to conclude
that a {\it pure imaginary} Immirzi parameter would shift $G$ for
+ and - modes (see~\cite{Randono} for closely related work). 
However this is where 
we lose connection with our work, because the + and -
modes are no longer R and L (which are real). We are currently working
on a suitable modification of the standard theory~\cite{future} that
does connect with the phenomenology in this Letter.

Should chirality be required to appear in the classical field
equations the implications could be even more dramatic. 
In~\cite{future} we shall demonstrate the following lemma.
Consider theories of gravity in $3+1$ dimensions which are, i) Functions of frame fields $e_\mu^a$ and a lorentzian connection $\omega_\mu^{ab}$ plus ordinary matter degrees of freedom. ii) Diffeomorphism invariant. ii) Invariant under local lorentz transformations. iii) The field equations expressed in terms of $e^a_\mu$ and $\omega_\mu^{ab}$ contain terms at most first order in derivatives.  These in general have parity asymmetric actions, nonetheless, the  linearization of the field equations around deSitter spacetime are those of 
general relativity (with $G_L =G_R$).
The implication is that if $TB$ were observed and this were due to a chiral effect in the linearized classical field equations,
then one of the very reasonable 
assumptions in this lemma would have to be violated.

{\it Conclusion} 
From the point of view of 
grand unification it would make much more sense if gravity were 
chiral~\cite{steph}. 
We have shown  that
should gravity be chiral at leading order in the linearized approximation, it would be easiest to 
detect gravitational waves precisely by making use of their chirality.
Current observations are not yet suitably sensitive, yet the future is bright.
But what other effects might gravitational chirality
have? One should bear in mind that the parameter
$\gamma$ could be dynamical, with chirality active during
inflation (when gravity waves were produced) but switched off
nowadays. $TB$ observations would then be the only way in which the
theory could be constrained.  If, however, gravitational chirality is
still present nowadays other interesting observational targets emerge,
which we mention in closing. The effect would appear in the quadrupole
formula for gravity wave emission, leading to different intensities
for L and R.  In the case of the millisecond pulsar by symmetry any
polarization bias in one direction would be matched by the reverse
bias in the opposite direction. The total power emitted would
therefore be the same, but a small ``rocket effect'' would be present.
The issue of chirality in a gravity wave background has also been
discussed in the context of direct gravitational wave
detection~\cite{gravpol}. Most existing experiments are polarization
myopic, but this could change in future experiments. Other effects on
the CMB should also be studied, in the context of specific quantum gravity
theories, such as the emergence of circular polarization.

{\it Acknowledgements} We'd like to thank Chris Isham, Andrew Jaffe,
Ettore Majorana, Marco Peloso, Simone Speziale and Toby Wiseman for
helpful comments.

\end{document}